# Terahertz photodetection in scalable single-layer-graphene and hexagonal boron nitride heterostructures

M. Asgari,[1] L. Viti,[1] O. Balci,[2] S. M. Shinde,[2] J. Zhang,[2] H. Ramezani,[2] S. Sharma,[2] A. Meersha,[2] G. Menichetti,[3] C. McAleese,[4] B. Conran,[4] X. Wang,[4] A. Tomadin,[3] A. C. Ferrari,[2] M. S. Vitiello[1*]

[1] NEST, CNR-Istituto Nanoscienze and Scuola Normale Superiore, Piazza San Silvestro 12, Pisa 56127, Italy
[2] Cambridge Graphene Centre, University of Cambridge, Cambridge CB3 0FA, UK
[3] Dipartimento di Fisica, Università di Pisa, Largo Bruno Pontecorvo 3, 56127 Pisa, Italy
[4] AIXTRON Ltd. Buckingway Business Park Anderson Rd, Swavesey, Cambridge CB24 4FQ, UK

[*]Corresponding author: miriam.vitiello@sns.it

**The unique optoelectronic properties of single layer graphene (SLG) are ideal for the development of photonic devices across a broad range of frequencies, from X-rays to microwaves. In the terahertz (THz) range (0.1-10 THz frequency) this has led to the development of optical modulators, non-linear sources, and photodetectors, with state-of-the-art performances. A key challenge is the integration of SLG-based active elements with pre-existing technological platforms in a scalable way, while maintaining performance level unperturbed. Here, we report on the development of room temperature THz detection in large-area SLG, grown by chemical vapor deposition (CVD), integrated in antenna-coupled field effect transistors. We selectively activate the photo-thermoelectric detection dynamics, and we employ different dielectric configurations on SLG on $Al_2O_3$ with and without large-area CVD hBN capping to investigate their effect on SLG thermoelectric properties underpinning photodetection. With these scalable architectures, response times ~5ns and noise equivalent powers ~1nWHz$^{-1/2}$ are achieved under zero-bias operation. This shows the feasibility of scalable, large-area, layered materials heterostructures for THz detection.**

Two-dimensional (2D) layered materials (LMs) and related heterostructures (LMHs) are a versatile platform for engineering optoelectronic and photonic devices[1,2,3]. They can be synthesised with wafer-scale methods[2,4,5] and stacked to form LMHs. Being compatible with Si and a wider range of III-V materials and substrates[2,5-7] they open new prospects for emerging research domains such as future high-density optical communications[6], high-speed datacom[8], quantum nanophotonics[9] and optoelectronics[1,3]. In particular, LMs are a versatile platform for devising photodetectors (PDs) operating across a broad range of frequencies from microwaves[10], to telecom[8], visible[1-3], and X-rays[11].

Terahertz (THz) radiation ($\nu$=0.1-10 THz) finds application in biomedicine[12], security[13], spectroscopy[14], cultural heritage[15], astronomy[16], real-time imaging[17] and high data-rate communications[18]. In this frequency range, optoelectronic systems employing LMs realized in scalable processes have mostly been developed for light modulation[19,20] and nonlinear optics applications[21,22], whereas similar processes for nanoscale receivers are still at an early stage, and limited to few examples involving scalable large-area single layer graphene (SLG)[23-28].

The challenge is twofold: first, the quest for technological maturity requires integration with established platforms, such as complementary metal oxide semiconductors (CMOS); second, performance in SLG grown by chemical vapour deposition (CVD)[24-28] still do not match those obtained with exfoliated hBN/SLG/hBN heterostructures.[29-34] Above 1 THz, few ns response times ($\tau$) and noise equivalent power (NEP) ~1 nWHz$^{-1/2}$ have been reported at room-temperature (RT) on THz PDs based on large-area (~cm$^2$) CVD SLG[28], but these are still inferior to those reported in high-quality hBN-encapsulated SLG ($\tau$=880 ps, NEP=80 pWHz$^{-1/2}$)[29,31], in Schottky diodes[35] (NEP ~100 pWHz$^{-1/2}$, electrical bandwidth ~40 GHz, operation frequency <1.8 THz), in high-electron-mobility transistors[36] (HEMTs, NEP ~100 pWHz$^{-1/2}$, $\tau$<10 ps), or in portable THz cameras based on FETs or microbolometer arrays[37,38] (NEP~30 pWHz$^{-1/2}$, $\tau$>10 µs). Table I summarizes the performance of room-temperature THz detectors realized with scalable graphene-based structures.

**TABLE I**. Performance of room-temperature THz detectors based on scalable, large-area graphene.

| Material | Frequency Range (THz) | NEP (nWHz$^{-1/2}$) | Response time (ns) |
|---|---|---|---|
| Epitaxial graphene on SiC[23] | 0.2-0.4 | 80 | n.a. |
| CVD graphene[24] | 0.36 | 10 | n.a. |
| CVD graphene[25] | 0.4 | 0.13 | n.a. |
| CVD graphene[26] | 0.6 | ~0.5 | n.a. |
| CVD graphene[27] | ~2 | ~150 | n.a. |
| CVD graphene[28] | ~3 | ~1 (mean 4.3) | ~5 |
| CVD graphene, CVD hBN heterostructure | ~3 | ~1 (mean 3.0) | ~7 |

As a step towards a wider commercial development of THz photodetectors based on LMs and LMHs, we propose here the use of substrate treatment,[39] or large-area encapsulation,[40] in order to reach stable and repeatable state-of-the-art performances at room-temperature. We develop RT photodetectors operating at 2.8 THz based on large-area (~1×1cm$^2$) CVD SLG, with and without large-area (~1×1cm$^2$) CVD hBN capping, Fig.1a.

A ~10 nm thick layer of $Al_2O_3$ is deposited on $SiO_2$/Si substrate (resistivity~10kΩ·cm) by atomic layer deposition (ALD) (see Supplementary Information). SLG is grown in a hot wall CVD system using ~30 μm thick Cu foil as substrate. The foil, suspended on a quartz holder and loaded into the CVD system, is annealed at 1050 °C for 2h under $H_2$ gas (100 sccm) at 760 Torr and cooled down to RT. For the growth, the foil is annealed at 1050 °C with 50 sccm hydrogen flow at 0.4 Torr for 2h. 5 sccm $CH_4$ is introduced to start growth, which is completed in 30mins by stopping the $CH_4$ flow. The system is then naturally cooled down to RT under 50 sccm $H_2$ flow. As-grown SLG/Cu is spin-coated with poly (methyl methacrylate) (PMMA) (A4-950) at 1000 rpm for 1 min and baked at 80°C for 10 mins. PMMA-coated SLG/Cu is kept in water overnight to oxidize Cu foil. PMMA/SLG is then electrochemically delaminated by applying 2 V between Pt anode and PMMA/SLG/Cu cathode in a NaOH aqueous electrolyte (~1M). The PMMA/SLG stack is cleaned in water and transferred on $Al_2O_3$/$SiO_2$/Si substrates, which are then baked at 80°C for 10mins after ~10 h natural drying. PMMA is removed by soaking in acetone and isopropyl alcohol (IPA).

hBN is grown on c-plane $Al_2O_3$ (0001) at 1400°C, 500 mbar for 30 minutes in an AIXTRON CCS 2D reactor. 10 sccm $N_2$ is used to transport the single-source precursor, borazine, to the reactor. Before hBN growth, the sapphire substrates are annealed in $H_2$ atmosphere for 5 mins, at 750 mbar and 1180°C. As-grown hBN on *c*-plane sapphire is then spin-coated with PMMA (A4-950) at 1000 rpm for 1min and baked at 80 °C for 10 mins. PMMA-coated hBN on sapphire is kept in ~8% $H_3PO_4$ for ~10 h to delaminate PMMA/hBN. This is cleaned in water and transferred on SLG/$Al_2O_3$/$SiO_2$/Si. After natural drying, this is baked at 80 °C for 10 mins. PMMA is removed by soaking it in acetone and IPA.

As-grown and transferred SLG and hBN are characterized by Raman spectroscopy with a Renishaw InVia spectrometer equipped with 100× objective at 514.5 nm. A statistical analysis of 7 spectra on as-grown SLG on Cu, 27 spectra on SLG on $Al_2O_3$/$SiO_2$/Si, 36 spectra on hBN/$Al_2O_3$/$SiO_2$/Si and 26 spectra on hBN/SLG/$Al_2O_3$/$SiO_2$/Si is performed to estimate defect density and doping. Errors are calculated from the standard deviation across different spectra, the spectrometer resolution (~1cm$^{-1}$) and the uncertainty associated with the different methods to estimate doping from the position of G peak Pos(G), its full-width-half-maximum FWHM(G), the intensity and area ratios I(2D)/I(G), A(2D)/A(G), and the position of 2D peak Pos(2D).[41-43] The Raman spectrum of as-grown SLG on Cu is in Fig.1b, after Cu photoluminescence removal.[44] The 2D peak is a single Lorentzian with FWHM(2D)=30±5 cm$^{-1}$, which is a signature of SLG.[45] Pos(G)=1584±2 cm$^{-1}$, FWHM(G)=16±4 cm$^{-1}$, Pos(2D)=2698±2 cm$^{-1}$, I(2D)/I(G)=3.8±1.0 A(2D)/A(G)=7.2±1.8. No D peak is observed, indicating negligible density of Raman active defects.[46,47] A prototypical Raman spectrum of SLG/Al$_2$O$_3$/SiO$_2$/Si is in Fig.1b. The 2D peak retains its single-Lorentzian line shape with FWHM(2D)=36±2 cm$^{-1}$. Pos(G)=1597±3 cm$^{-1}$, FWHM(G)=14±2 cm$^{-1}$, Pos(2D)=2692±2 cm$^{-1}$, I(2D)/I(G)=1.9±0.5 and A(2D)/A(G)=4.9±0.5, indicating a p-doping with Fermi level E$_F$=290±90 meV,[41,42] which corresponds to a carrier concentration n=6.5±3.3×10$^{12}$ cm$^{-2}$.[41,42] I(D)/I(G)=0.03±0.04 corresponds to a defect density n$_D$=1.7±0.7×10$^{10}$ cm$^{-2}$ for excitation energy 2.41eV and E$_F$=290±90 meV.[43] A Raman spectrum of the transferred hBN on Al$_2$O$_3$/SiO$_2$/Si is in Fig.1b. The position of E$_{2g}$ peak, Pos(E$_{2g}$)=1371±1 cm$^{-1}$, with FWHM(E$_{2g}$)=26±1 cm$^{-1}$, indicating that it is hexagonal BN.[48] A Raman spectrum of SLG capped by hBN is in Fig.1b. The single-Lorentzian 2D peak has FWHM(2D)=37±2 cm$^{-1}$. Pos(G)=1598±3 cm$^{-1}$, FWHM(G)=13±1 cm$^{-1}$, Pos(2D)=2692±2 cm$^{-1}$, I(2D)/I(G)=1.8±0.3 and A(2D)/A(G)=5.1±0.6, indicating a p-doping with E$_F$=300±100 meV,[41,42] which corresponds to n=7.2±4.1×10$^{12}$ cm$^{-2}$.[41,42]

Graphene field-effect transistors (GFETs) are fabricated by means of electron beam lithography (EBL) (Zeiss UltraPlus), ALD (Oxford, OpAL) and metal evaporation. We define U-shaped channels with length L$_c$= 2400 nm and width W$_c$= 1500 nm through reactive ion etching (O$_2$ for uncapped, CF$_4$/O$_2$ mixture for hBN-capped samples). We pattern source (s) and drain (d) contacts via EBL, and fabrication is finalized by depositing a 40nm thick Pd layer by thermal evaporation and lift-off. For hBN-capped devices, this results in edge-contacts to the SLG channel, showing contact resistance R$_0$~1 kΩ, similar to the case of uncapped samples. We then define the top-gate oxide (~40 nm HfO$_2$, grown by ALD) above the channel, and finalize the fabrication by depositing 5/90 nm Cr/Au to establish the top-gate (g) electrodes. To reduce the detector shunt capacitance, s and d electrodes are connected to a coplanar strip-line.[31] A schematic cross-section of the fabricated devices is in Fig.1a. For each material combination, Al$_2$O$_3$/SLG/HfO$_2$ and Al$_2$O$_3$/SLG/hBN/HfO$_2$, two distinct architectures are conceived, devised and investigated. The first is based on single-top-gated GFETs integrated with a planar bow-tie antenna, with radius 24 μm and flare angle 90° (Fig.1c,d). The antenna arms are connected to the s and g electrodes. The second design features two top-gates, connected to the left and right arms of a 24 μm radius bow-tie antenna. The antenna dimensions are chosen to be resonant with a radiation frequency of 2.8 THz.[28]

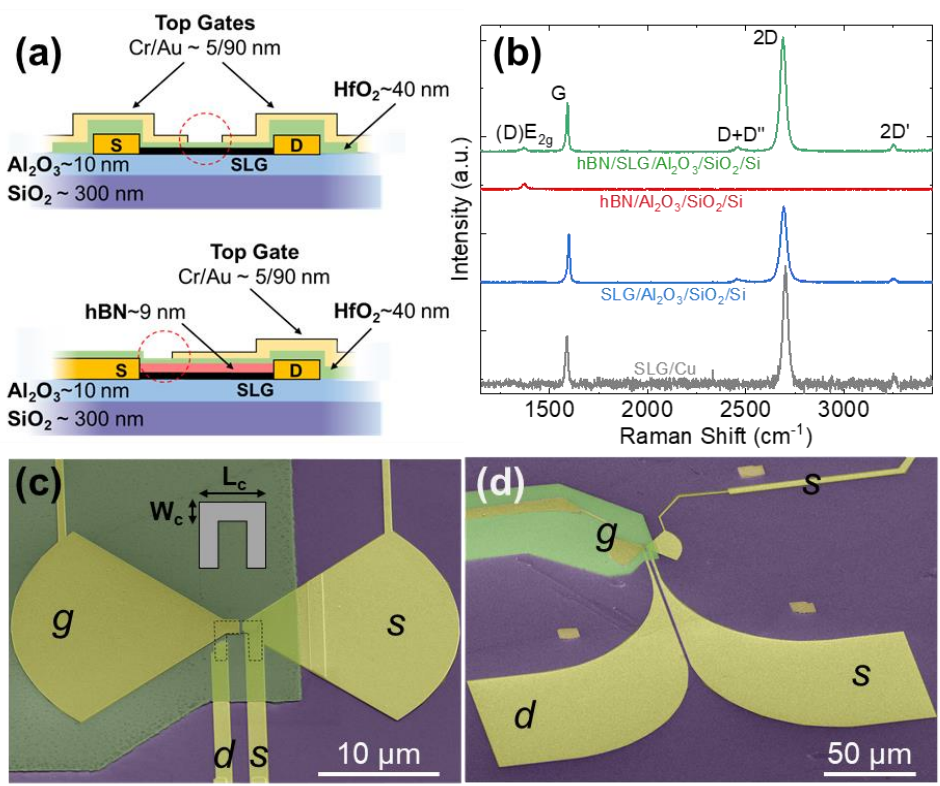

**Figure 1.** (a) Schematic cross-section of the GFETs. Top: double-gated Al$_2$O$_3$/SLG/HfO$_2$. Bottom: single-gated Al$_2$O$_3$/SLG/hBN/HfO$_2$. Dashed red circles indicate the position of the THz-induced field enhancement. (b) Raman spectra of as-grown SLG on Cu and SLG transferred on Al$_2$O$_3$ with and without hBN capping. (c,d) False colour scanning electron micrographs of a single-gated device. The inset shows the U-shaped channel.

These geometries are selected to activate the photo-thermoelectric (PTE) effect as dominant detection mechanism.[31,32,49] This requires a spatial asymmetry along the SD channel.[3] In single-gated systems, the asymmetry is established by the spatial gradient of the electronic temperature (T$_e$) in the SLG channel, induced by the absorption of THz light in the antenna gap, which is located close to the $s$ electrode.[31] Instead, in $p$-$n$ junctions, the electronic distribution is symmetrically heated at the centre of the $s$-$d$ channel, where the antenna gap is located.[29] Here, the asymmetry is determined by the longitudinal variation of the SLG Seebeck coefficient (S$_b$), whose profile along the $s$-$d$ direction can be electrostatically defined by applying distinct gate voltages (V$_g$) on the left and right sides of the junction.

We activate a dominant PTE by design, since SLG displays unique thermoelectric properties, due to its low electronic specific heat (~2000 $k_B$μm$^{-2}$ at RT,[50] where $k_B$ is the Boltzmann constant), ultrafast (~30 fs) carrier thermalization dynamics[51] and slower (~4 ps) electron-phonon cooling.[52,53] Thus, it offers an outstanding route for performance optimization by means of e.g., carrier lifetime engineering[49] or coupling to plasmonic-polaritonic quasi-particles.[54] PTE also allows for broadband,[49] room-temperature, zero-bias operation.[8]

We first characterize the devices electrically, by measuring the source-drain current ($I_{sd}$) as a function of $V_g$. The resistance (R) curve for a single-gated device is in Fig.2a. Instead, Fig.2b shows the resistance map of a *p-n* junction device, measured as a function of the left- and right-gate voltages ($V_{gL}$ and $V_{gR}$). We extract the field-effect mobility ($\mu_{FE}$) for electrons and holes and the residual carrier density ($n_0$), by using the fitting function[55] R= $R_0+(L_c/W_c)\cdot(1/n_{2d}e\mu_{FE})$, where $n_{2d}=[n_0^2 + (C_g/e\cdot(V_g-V_{CNP}))^2]^{1/2}$ is the carrier density,[55] $C_g$ is the gate-capacitance per unit area and $V_{CNP}$ is the charge neutrality point. We get $\mu_{FE}$=300-2000 cm$^2$V$^{-1}$s$^{-1}$ for Al$_2$O$_3$/SLG-based GFETs and $\mu_{FE}$=3000-9000 cm$^2$V$^{-1}$s$^{-1}$ for hBN-capped ones.

Fig.2c plots $n_0$ as a function of $\mu$ for the complete batch of devices, where $\mu$ is the average field-effect mobility of electrons and holes: each GFET is represented by a coloured dot. From this comparison, hBN-capped samples show lower $n_0$ and higher $\mu$ with respect to Al$_2$O$_3$/SLG/HfO$_2$ devices. The field-effect measurements (Figs.2a,b) can be used to evaluate $S_b$, which determines the PTE response of SLG-based PDs. This can be done starting from the experimental conductivity ($\sigma$) and using the Mott equation,[30] which, however, is not accurate at low carrier densities,[56] i.e. close to $V_{CNP}$. Thus, we theoretically calculate $S_b$ using an effective medium theory (EMT)[56] in the framework of the linear Boltzmann equation[53,57] (see Supplementary Information). Fig.2d shows the results of the EMT model applied to the device of Fig.2a, for which we get a maximum thermopower $S_{max}$~70 $\mu$VK$^{-1}$.

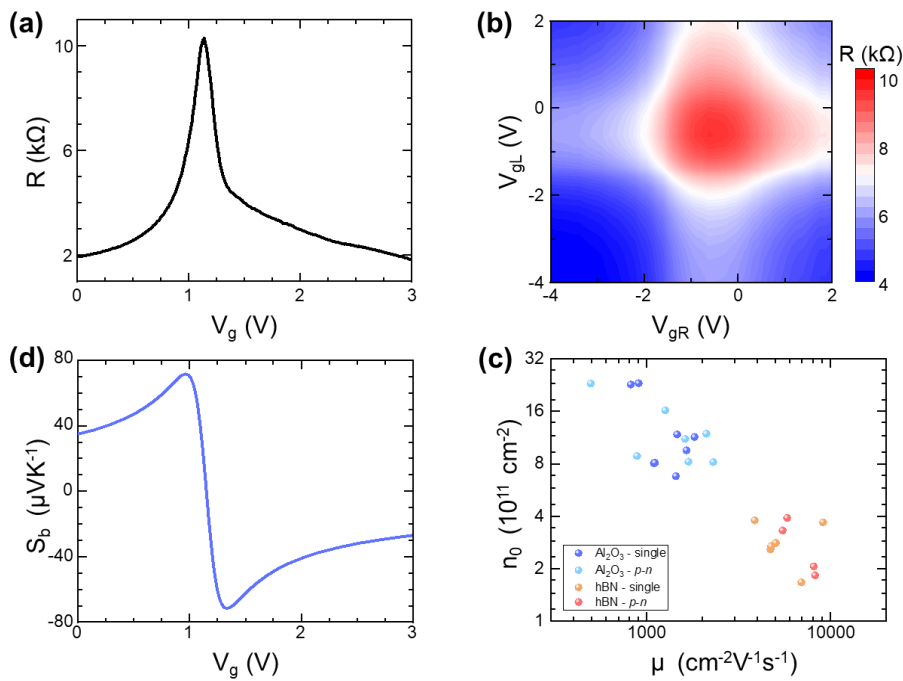

**Figure 2.** (a) Channel resistance *vs*. $V_g$ for an hBN-capped, single-gated GFET. (b) Map of R as a function of $V_{gR}$ and $V_{gL}$ for a hBN-capped *p-n* junction. (c) Chart of $n_0$ *vs*. $\mu_{FE}$ for all the measured devices. Blue (light-blue) dots represent single-gated (double-gated) FETs without hBN-capping. Orange (red) dots represent hBN-capped single-gated (double-gated) FETs. (d) $S_b$ calculated with the Boltzmann EMT[51,56,57] for the R curve in (a).

We then evaluate the PDs optical figures of merit: voltage responsivity ($R_v$), NEP and τ. The detectors are illuminated by a 2.8 THz quantum cascade laser (QCL), driven in pulsed mode (repetition rate 40 kHz, duty cycle 4%) delivering a peak power ~25 mW, corresponding to an average power ~1 mW. The antenna axis is oriented parallel to the linearly polarized electric field. The beam is focused by two TPX lenses onto a ~200 μm radius circular spot. We select an intermediate average power $P_0$=0.4 mW to characterize the PDs, to avoid QCL overheating. The corresponding average intensity in the focal point is $I_0$=0.32 Wcm$^{-2}$.

We measure the photovoltage (Δu) at the *d* electrode, while keeping *s* grounded. Δu is amplified by a voltage preamplifier (DL Instruments, M1201, gain γ=1000) and sent to a lock-in (Stanford Research, 5210). We use a square-wave envelope with frequency $f_{mod}$=1.333 kHz as lock-in reference and as triggering signal for the QCL pulse trains. Δu can be inferred from the demodulated lock-in signal ($V_{LI}$) via the relation[30] Δu=(π√2/2)$V_{LI}$/γ, where the pre-factor π√2/2 takes into account that the lock-in measures the root mean square of the fundamental Fourier component of the square wave produced by the QCL modulation. $R_v$ is calculated from the ratio between Δu and the power $P_a$=$I_0 A_{eff}$ impinging on the detector, with $A_{eff}$ the detector effective area, assumed equal to the diffraction limited area[30] $A_{eff}$=$λ^2$/4=2800 μm$^2$, where λ is the free-space wavelength. We calculate the curve of $R_v$ *vs*. $V_g$ for each single-gated GFET and the map of $R_v$ *vs*. $V_{gL}$ and $V_{gR}$ for each *p-n* junction. Typical examples of $R_v$ *vs*. $V_g$ plots are in Fig.3a,b for a single-gated GFET and a *p-n* junction, respectively. The photoresponse in single-gated GFETs follows the profile of $S_b$, with an offset $S_{bu}$≅$S_b$ ($V_g$=0 V), with PTE voltage[30] $V_{PTE}$= Δ$T_e$·($S_b$ - $S_{bu}$), where Δ$T_e$ is the $T_e$ gradient between *s* (hot) and *d* (cold) sides of the SLG channel, and $S_{bu}$ the Seebeck coefficient of the ungated region between the *s* and *g* electrodes. Fig.3a compares the measured $R_v$ *vs*. $V_g$ and the theoretical PTE responsivity $R_{PTE}$ *vs*. $V_g$, with $R_{PTE}$ inferred from $V_{PTE}$ by considering Δ$T_e$/$P_0$~1.5 K/mW, showing good qualitative agreement between the two curves. The discrepancy between the theoretical and experimental responsivities at large positive $V_g$ is ascribed to the fact that the adopted theoretical model (see Supporting Information) only includes electron scattering with charged Coulomb impurities as dominant effect limiting the conductivity, possibly neglecting additional contributions, e.g. phonon scattering or carrier inhomogeneities at the contacts. A dominant PTE detection mechanism is also observed in *p-n* junctions. $E_F$ in SLG can be tuned across the Dirac point by the electrostatic gating applied to the left and right sides of the junction. The non-monotonic dependence of $S_b$ on $E_F$ leads to multiple sign changes in $R_v$, resulting in a six-fold pattern[29] in the $R_v$ map (Fig.3b), a distinctive feature of PTE.[26,29]

The sensitivity of THz detectors is evaluated through the NEP,[29] defined as the ratio between noise figure and responsivity. In order to calculate the NEP, it is important to give a rigorous evaluation of the noise spectral density (NSD).[26,29-31] Thus, we measure the GFETs NSD with a lock-in amplifier (Zurich Inst., UHFLI): the *s* electrode is grounded and the signal, demodulated by the lock-in, is collected at the *d* electrode, while a sweep of the modulation frequency is performed. The results are in Fig.4c for a 50 Ω test resistor and for a prototypical SLG-based device. The white noise floor for the 50Ω resistor is dominated by the lock-in noise figure. The Johnson-Nyquist NSD formula gives[30] $N_J=(4k_BTR)^{1/2}$=0.91 nVHz$^{-1/2}$ for a 50Ω resistor operated at RT, whereas our instrumental noise floor is ~8 nVHz$^{-1/2}$, as expected for the noise level of the employed lock-in.[58] The NSD of one of the GFETs (R=9 kΩ in Fig.4c) is dominated by the $1/f$ component[59] for modulation frequency <1kHz and flattens at NSD <14 nVHz$^{-1/2}$ at higher frequencies, in agreement with the theoretically expected $N_J$=12.3 nVHz$^{-1/2}$. The GFET $N_J$ is thus the main contribution to the overall noise figure in our setup (with pre-amplifier NSD~7 nVHz$^{-1/2}$). The measured NSD at 1.333kHz is then used to calculate NEP (Fig.3d,e) as a function of the voltages applied to the gate electrodes.

We then characterize the detection speed by recording the time trace of Δu with an oscilloscope (Tektronix DPO520-4B, bandwidth 2 GHz). We use a THz pulse duration ~1.6 μs and we amplify the PD output with a high-bandwidth (1.1 GHz) voltage preamplifier (Femto, DUPVA-1-70) before the oscilloscope. We drive the QCL into the negative differential resistance regime,[30] which results in electronic instabilities that correspond to an intermittent output power: the QCL undergoes intensity fluctuations with characteristic time constants $\tau_{qcl}$~0.9 ns. This strategy allows us to test the bandwidth of our PDs up to a maximum $(2\pi\tau_{qcl})^{-1}$=180 MHz. Fig.3f shows the waveform recorded by a single-gated GFET during an intensity fluctuation of the pulsed QCL. We evaluate τ from exponential fits to the waveform (see Supplementary Information). We get τ=7-20 ns, with a mean value ~12 ns, corresponding to a bandwidth ~15 MHz.

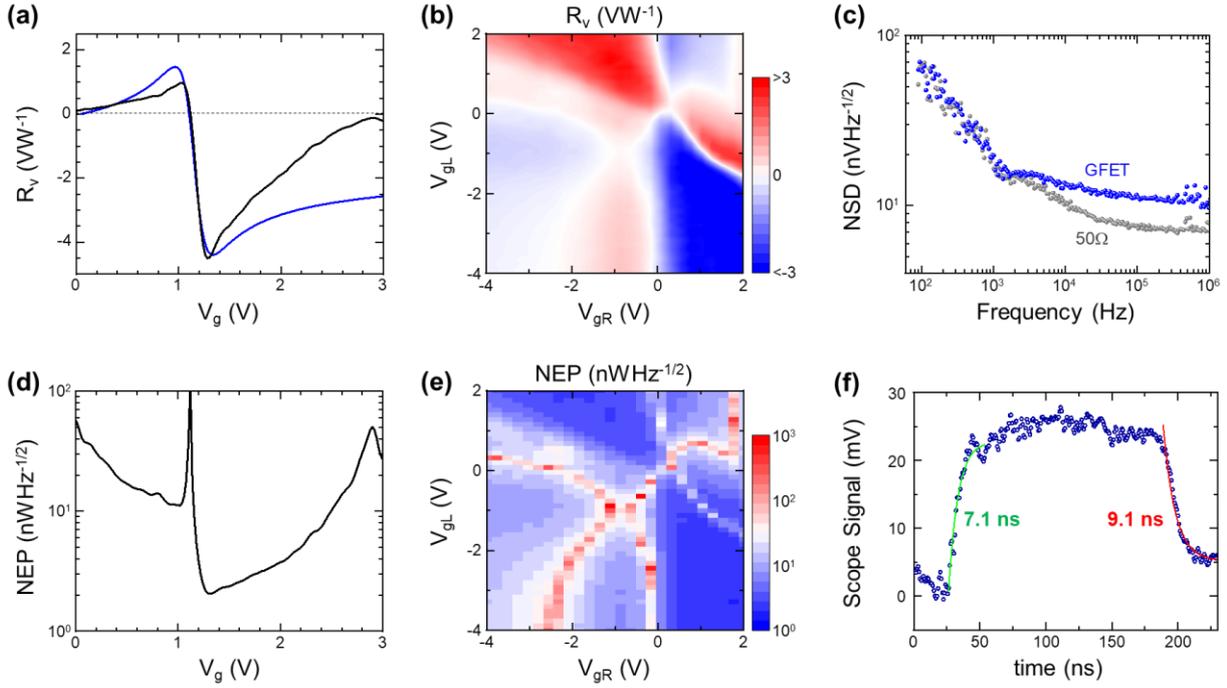

**Figure 3**. (a) $R_v$ as a function of $V_g$ for an hBN-capped single-gated GFET. The experimental curve (black line) is compared with the theoretical PTE response (blue line), evaluated by EMT. (b) $R_v$ map of an hBN-capped *p-n* junction, as a function of $V_{gL}$ and $V_{gR}$. (c) NSD of a GFET and of a 50 Ω resistor, measured by sweeping the reference frequency of the lock-in from 100 Hz to 1 MHz. (d,e) NEP of a single-gated and a *p-n* junction GFETs as a function of gate voltage(s). (f) Time trace of an intensity fluctuation of the QCL. The rising and falling edges are fitted with exponential functions to retrieve τ.

Statistical analysis is applied to 28 devices to evaluate performance variability and identify correlations between electrical and optical properties. We first consider NEP variability. For $Al_2O_3$/SLG/$HfO_2$ devices, we get a mean value ~7.6 nWHz$^{-½}$ and an interquartile range[28] (IQR) ~4.0 nWHz$^{-½}$. For hBN-capped PDs, we have mean NEP~3.0 nWHz$^{-½}$ with IQR~1.4 nWHz$^{-½}$, which represents a variability improvement of a factor >2 with respect to $SiO_2$/SLG[24] and $Al_2O_3$/SLG PDs. We then evaluate correlations between NEP, $S_b$ and $n_0$ using the Pearson coefficient[60] (ρ) as a metric. ρ($v_1$,$v_2$) represents the measure of linear correlation between two discrete variables $v_1$ and $v_2$: |ρ|=1 indicates an exact linear dependence and ρ=0 indicates no linear correlation. We get ρ($S_{max}$,$n_0$)= −0.95 for both hBN-capped and uncapped architectures, where $S_{max}$ is the maximum |$S_b$| in the investigated $V_g$ range, calculated with the EMT model. The scatter plot of $S_{max}$ *vs.* $n_0$ in Fig.4a shows that hBN/SLG/$Al_2O_3$ LMHs have slightly larger $S_{max}$, even though they have significantly smaller $n_0$. This is due to the different dielectric environment: the larger $\varepsilon_r$ of $HfO_2$ (with respect to hBN) on top of SLG is beneficial in terms of thermopower.[56] This similarity in $S_{max}$ is reflected in the detectors NEP, where the difference between the two material architectures is not as pronounced as the difference in $n_0$ (Fig.4b). However, in agreement with results obtained on $SiO_2$/SLG/$HfO_2$ heterostructures,[28] NEP increases for larger $n_0$: $\rho(\log(NEP),\log(n_0))=0.4$. These correlations confirm that the physical mechanism underpinning THz detection is, as expected, PTE.

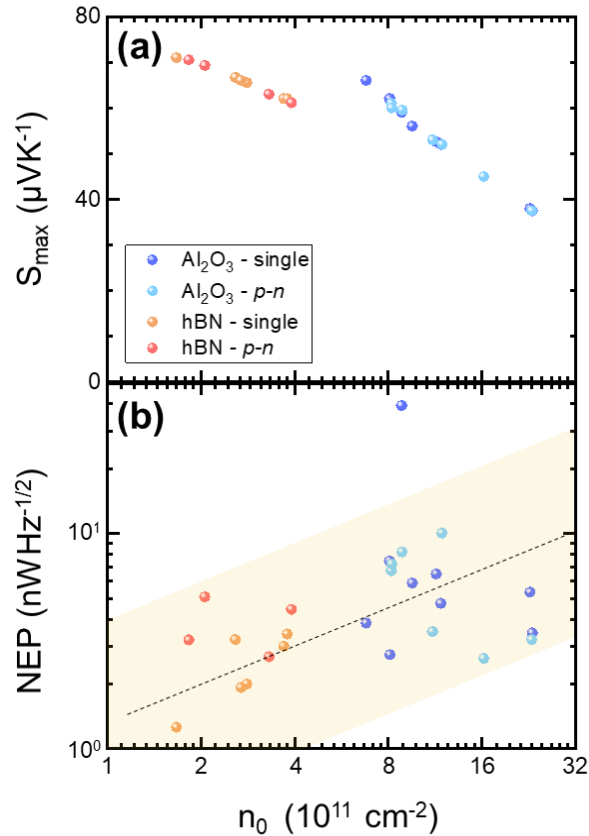

**Figure 4.** (a) Scatter plot of $S_{max}$ *vs*. $n_0$. Differently coloured dots identify different material/geometry combinations. $S_{max}$ and $n_0$ have negative correlation. (b) NEP *vs*. $n_0$ chart, showing positive correlation: $\rho(\log(NEP),\log(n_0))=0.4$. The dotted line is a guide for the eye.

Thus, the $Al_2O_3$ termination alone does not show a significant performance improvement over $SiO_2$/Si substrates,[28] whereas large-area $HfO_2$/hBN/SLG/$Al_2O_3$ LMHs present advantages both in terms of absolute optical performance (average NEP~3.0 nWHz$^{-½}$) and performance variability (IQR~1.40 nWHz$^{-½}$). It is worth mentioning that large area hBN-top-encapsulation significantly reduces the device performance variability by more than a factor 2 with respect to Ref. 28.

In summary, we report on THz PDs realized with large-area graphene and large-area hBN in wafer-scale compliant processes, capable of mitigating material degradation with respect to the quality benchmark of hBN-encapsulated SLG.[29-31] We demonstrate THz detection in a layered material heterostructure obtained by consecutive transfer of CVD graphene and CVD hexagonal boron nitride, a fabrication technique that is fully compatible with standard CMOS processing. This makes our PDs suited for real-time imaging and short-range (~10 m) THz communication applications, enabling multi-pixel architectures. A further benefit can come from the full large-area encapsulation of SLG in CVD-based hBN/SLG/hBN heterostructures.

**Supplementary material**

See Supplementary material for description of: $Al_2O_3$ ALD, thermopower calculation and detection speed analysis.

**Acknowledgements**

We acknowledge funding from ERC Projects 681379 (SPRINT), Hetero2D, GSYNCOR, the EU Graphene and Quantum Flagships, the Marie Curie H2020-MSCA-ITN2017 TeraApps (765426) grant, the CNR project (TEROCODE), and EPSRC EPSRC Grants EP/L016087/1, EP/K01711X/1, EP/K017144/1, EP/N010345/1, EP/V000055/1, DSTL. G.M. acknowledges support from the University of Pisa under the "PRA - Progetti di Ricerca di Ateneo" (Institutional Research Grants) - Project No. PRA 2020-2021 92. For the purpose of open access, the authors applied a Creative Commons Attribution (CC BY) license to any Author Accepted Manuscript version arising from this submission.

**Data Availability Statement**

The data that support the findings of this study are available from the corresponding author upon reasonable request.